\magnification \magstep1
\raggedbottom
\openup 2\jot 
\voffset6truemm
\def\cstok#1{\leavevmode\thinspace\hbox{\vrule\vtop{\vbox{\hrule\kern1pt
\hbox{\vphantom{\tt/}\thinspace{\tt#1}\thinspace}}
\kern1pt\hrule}\vrule}\thinspace}
\centerline {\bf ON THE GREEN FUNCTIONS OF}
\centerline {\bf GRAVITATIONAL RADIATION THEORY}
\vskip 1cm
\leftline {Giampiero Esposito}
\vskip 1cm
\noindent
{\it Istituto Nazionale di Fisica Nucleare, Sezione
di Napoli, Complesso Universitario di Monte S. Angelo, Via Cintia,
Edificio N', 80126 Napoli, Italy}
\vskip 0.3cm
\noindent
{\it Universit\`a di Napoli Federico II, Dipartimento
di Scienze Fisiche, Complesso Universitario di Monte S. Angelo,
Via Cintia, Edificio N', 80126 Napoli, Italy}
\vskip 1cm
\noindent
{\bf Abstract.}
Previous work in the literature has studied gravitational radiation
in black-hole collisions at the speed of light. In particular, it had
been proved that the perturbative field equations may all be reduced to
equations in only two independent variables, by virtue of a conformal
symmetry at each order in perturbation theory. The Green function for the
perturbative field equations is here analyzed by studying the corresponding
second-order hyperbolic operator with variable coefficients, instead of
using the reduction method from the retarded flat-space Green function
in four dimensions. After reduction to canonical form of this hyperbolic
operator, the integral representation of the solution in terms of the
Riemann function is obtained. The Riemann function solves a characteristic
initial-value problem for which analytic formulae leading to the
numerical solution are derived.
\vskip 100cm
\leftline {\bf 1. Introduction}
\vskip 0.3cm
\noindent
The construction of suitable inverses of differential operators lies still
at the very heart of many profound properties in classical and quantum
field theory. For example, the theory of small disturbances in local field
theory can only be built if suitable invertible operators are considered [1].
In a path-integral formulation, these correspond to the gauge-field and
ghost operators, respectively [2, 3]. Moreover, the Peierls bracket on the
space of physical observables, which is a Poisson bracket preserving the
invariance under the full infinite-dimensional symmetry group of the theory,
is obtained from the advanced and retarded Green functions of the theory via 
the supercommutator function [1--4], and leads possibly to a deeper approach
to quantization. Last, but not least, a perturbation approach to classical
general relativity relies heavily on a careful construction of Green functions
of operators of hyperbolic [5--8] and elliptic [9, 10] type. In particular,
following [5--8], we shall be concerned with the axisymmetric collision of two
black holes travelling at the speed of light, each described in the
centre-of-mass frame before the collision by an impulsive plane-fronted shock
wave with energy $\mu$. One then passes to a new frame to which a large Lorentz
boost is applied. There the energy $\nu=\mu e^{\alpha}$ of the incoming shock
$1$ obeys $\nu >> \lambda$, where $\lambda=\mu e^{-\alpha}$ is the energy
of the incoming shock $2$ and ${\rm e}^{\alpha} \equiv \sqrt{{1+\beta}\over
{1-\beta}}$ ($\beta$ being the usual relativistic parameter). 
In the boosted frame, to the future of the strong
shock $1$, the metric can be expanded in the form [6, 8]
$$
g_{ab} \sim \nu^{2} \left[\eta_{ab}+\sum_{i=1}^{\infty}
\left({\lambda \over \nu}\right)^{i}h_{ab}^{(i)}\right]
\eqno (1.1)
$$
where $\eta_{ab}$ is the standard notation for the Minkowski metric. 
The task of solving the Einstein field equations becomes then a problem
in singular perturbation theory, having to find $h_{ab}^{(1)},h_{ab}^{(2)},...$
by solving the linearized field equations at first, second, ... order 
respectively in ${\lambda \over \nu}$, once that characteristic initial
data are given just to the future of the strong shock $1$. The perturbation
series (1.1) is physically relevant because, on boosting back to the
centre-of-mass frame, it is found to give an accurate description of
space-time geometry where gravitational radiation propagates at small 
angles away from the forward symmetry axis ${\hat \theta}=0$. The news
function $c_{0}$, which describes gravitational radiation arriving at
future null infinity in the centre-of-mass frame, is expected to have the
convergent series expansion [6, 8]
$$
c_{0}({\hat \tau},{\hat \theta})=\sum_{n=0}^{\infty}
a_{2n}({\hat \tau}/\mu)(\sin {\hat \theta})^{2n}
\eqno (1.2)
$$
with $\hat \tau$ a suitable retarded time coordinate, and $\mu$ the energy
of each incoming black hole in the centre-of-mass frame. In [6, 8] a very
useful analytic expression of $a_{2}({\hat \tau}/\mu)$ was derived,
exploiting the property that perturbative field equations may all be reduced
to equations in only two independent variables, by virtue of a remarkable
conformal symmetry at each order in perturbation theory. The Green 
function for perturbative field equations was then found by reduction 
from the retarded flat-space Green function in four dimensions. 

However, a {\it direct} approach to the evaluation of Green functions
appears both desirable and helpful in general, and it has been our 
aim to pursue such a line of investigation. For this purpose, reduction
to two dimensions with the associated hyperbolic operator is studied
again in section 2. Section 3 performs reduction to canonical form
with the associated Riemann function. Equations for the Goursat problem
obeyed by the Riemann function are derived in section 4, while the
corresponding numerical algorithm is discussed in section 5.
\vskip 0.3cm
\leftline {\bf 2. Reduction to two dimensions and the associated operator}
\vskip 0.3cm
\noindent
As is well known from the work in [6] and [8], the field equations for
the first-order correction $h_{ab}^{(1)}$ in the expansion (1.1) are
particular cases of the general system given by the flat-space wave
equation (here $u \equiv {1\over \sqrt{2}}(z+t),
v \equiv {1\over \sqrt{2}}(z-t)$)
$$
\cstok{\ }\psi=2{\partial^{2}\psi \over \partial u \partial v}
+{1\over \rho}{\partial \over \partial \rho} \left(
\rho {\partial \psi \over \partial \rho}\right)
+{1\over \rho^{2}}{\partial^{2}\psi \over \partial \phi^{2}}=0
\eqno (2.1)
$$
supplemented by the boundary condition
$$
\psi(u=0)={\rm e}^{im \phi}\rho^{-n}f[8 \log(v \rho)-\sqrt{2}v]
\eqno (2.2a)
$$
$$
f(x)=0 \; \; \forall x < 0.
\eqno (2.2b)
$$
Moreover, $\psi$ should be of the form 
${\rm e}^{i m \phi}\rho^{-n}\chi(q,r)$
for $u \geq 0$, where
$$
q \equiv u \rho^{-2}
\eqno (2.3)
$$
$$
r \equiv 8 \log(\nu \rho)-\sqrt{2}v.
\eqno (2.4)
$$
For the homogeneous wave equation (2.1) there is no advantage in eliminating
$\rho$ and $\phi$ from the differential equation. However, the higher-order
metric perturbations turn out to obey inhomogeneous flat-space wave equations
of the form
$$
\cstok{\ }\psi=S
\eqno (2.5)
$$
where $S$ is a source term equal to ${\rm e}^{im \phi}\rho^{-(n+2)}H(q,r)$. This
leads to the following equation for 
$\chi \equiv {\rm e}^{-im \phi}\rho^{n}\psi$:
$$
{\cal L}_{m,n}\chi(q,r)=H(q,r)
\eqno (2.6)
$$
where ${\cal L}_{m,n}$ is an hyperbolic operator in the independent 
variables $q$ and $r$, and takes the form [6, 8]
$$ \eqalignno{
{\cal L}_{m,n}&=-(2\sqrt{2}+32q){\partial^{2}\over \partial q \partial r}
+4q^{2}{\partial^{2}\over \partial q^{2}}
+64{\partial^{2}\over \partial r^{2}} \cr
&+4(n+1)q{\partial \over \partial q}-16n{\partial \over \partial r}
+n^{2}-m^{2}.
&(2.7)\cr}
$$
The proof of hyperbolicity of ${\cal L}_{m,n}$, with the associated
normal hyperbolic form, can be found in 
section 3 of [6], and in [8]. The advantage
of studying Eq. (2.6) is twofold: to evaluate the solution at some
space-time point one has simply to integrate the product of $H$ and the
Green function $G_{m,n}$ of ${\cal L}_{m,n}$:
$$
\chi(q,r)=\int G_{m,n}(q,r;q_{0},r_{0})H(q_{0},r_{0})dq_{0}dr_{0}
\eqno (2.8)
$$
and the resulting numerical calculation of the solution is now 
feasible [7, 8].

Since we are interested in a direct approach to the evaluation of the
Green function $G_{m,n}$ in the $(q,r)$ coordinates, we begin by
noticing that, in all derivatives which are not mixed, the operator
${\cal L}_{m,n}$ can be made a constant coefficient operator 
upon setting
$$
\alpha \equiv \log(q)
\eqno (2.9)
$$
which implies
$$
q{\partial \over \partial q}={\partial \over \partial \alpha}
\eqno (2.10)
$$
$$
q^{2}{\partial^{2}\over \partial q^{2}}
={\partial^{2}\over \partial \alpha^{2}}
-{\partial \over \partial \alpha}
\eqno (2.11)
$$
and hence
$$ \eqalignno{
{\cal L}_{m,n}&=-(2\sqrt{2}{\rm e}^{-\alpha}+32)
{\partial^{2}\over \partial \alpha \partial r}
+4{\partial^{2}\over \partial \alpha^{2}}
+64{\partial^{2}\over \partial r^{2}} \cr
&+4n{\partial \over \partial \alpha}
-16n{\partial \over \partial r}+n^{2}-m^{2}.
&(2.12)\cr}
$$
This operator is further simplified upon defining the variable
$$
R \equiv {r\over 4}
\eqno (2.13)
$$
which implies that
$$ \eqalignno{
{\cal L}_{m,n}&=-\left({1\over \sqrt{2}}{\rm e}^{-\alpha}+8 \right)
{\partial^{2}\over \partial \alpha \partial R}
+4 \left({\partial^{2}\over \partial \alpha^{2}}
+{\partial^{2}\over \partial R^{2}}\right)\cr
&+4n \left({\partial \over \partial \alpha}
-{\partial \over \partial R}\right)+n^{2}-m^{2}.
&(2.14)\cr}
$$
This suggests defining yet new variables
$$
X \equiv \alpha+R
\eqno (2.15)
$$
$$
Y \equiv \alpha-R
\eqno (2.16)
$$
so that 
$$
{\partial^{2}\over \partial \alpha \partial R}
={\partial^{2}\over \partial X^{2}}-{\partial^{2}\over \partial Y^{2}}
\eqno (2.17)
$$
$$
{\partial^{2}\over \partial \alpha^{2}}
+{\partial^{2}\over \partial R^{2}}=2 \left(
{\partial^{2}\over \partial X^{2}}
+{\partial^{2}\over \partial Y^{2}}\right)
\eqno (2.18)
$$
$$
{\partial \over \partial \alpha}
-{\partial \over \partial R}=2{\partial \over \partial Y}
\eqno (2.19)
$$
and hence ${\cal L}_{m,n}$ reads eventually
$$ \eqalignno{
T_{m,n}&=16{\partial^{2}\over \partial Y^{2}}
+8n{\partial \over \partial Y}+n^{2}-m^{2} \cr
&-{1\over \sqrt{2}}{\rm e}^{-(X+Y)/2}
\left({\partial^{2}\over \partial X^{2}}
-{\partial^{2}\over \partial Y^{2}}\right)
&(2.20)\cr} 
$$
with Green function satisfying the equation
$$ \eqalignno{
\; & T_{m,n}G_{m,n}\left({\rm e}^{{X+Y}\over 2},2(X-Y);
{\rm e}^{{X_{0}+Y_{0}}\over 2},2(X_{0}-Y_{0}) \right) \cr
&={1\over 2}\delta \left({\rm e}^{{X+Y}\over 2}
-{\rm e}^{{X_{0}+Y_{0}}\over 2}\right)
\delta((X-Y)-(X_{0}-Y_{0})).
&(2.21)\cr}
$$
The operator $T_{m,n}$ is the sum of an elliptic operator in the
$Y$ variable and a two-dimensional wave operator `weighted' with the
exponential ${\rm e}^{-(X+Y)/2}$, which is the main source of technical
complications in these variables.
\vskip 0.3cm
\leftline {\bf 3. Reduction to canonical form and the Riemann function}
\vskip 0.3cm
\noindent
It is therefore more convenient, in our general analysis, to reduce first
Eq. (2.6) to canonical form, and then find an integral representation 
of the solution. Reduction to canonical form means that new coordinates
$x=x(q,r)$ and $y=y(q,r)$ are introduced such that the coefficients
of ${\partial^{2}\over \partial x^{2}}$ and 
${\partial^{2}\over \partial y^{2}}$ vanish. As is shown in [6, 8], 
this is achieved if
$$
{\partial x \over \partial r}={\partial y \over \partial r}=1
\eqno (3.1)
$$
$$
{\partial x \over \partial q}={1+8q \sqrt{2}+\sqrt{1+16q\sqrt{2}} \over
2\sqrt{2}q^{2}}
\eqno (3.2a)
$$
$$
{\partial y \over \partial q}={1+8q \sqrt{2}-\sqrt{1+16q\sqrt{2}} \over
2\sqrt{2}q^{2}}.
\eqno (3.3a)
$$
The resulting formulae are considerably simplified if one defines
$$
t \equiv \sqrt{1+16q \sqrt{2}}=t(x,y).
\eqno (3.4)
$$
The dependence of $t$ on $x$ and $y$ is obtained implicitly by 
solving the system [6, 8]
$$
x=r+ 8 \log \left({{t-1}\over 2}\right)-{8\over (t-1)}-4
\eqno (3.5)
$$
$$
y=r+ 8 \log \left({{t+1}\over 2}\right)+{8\over (t+1)}-4.
\eqno (3.6)
$$
This leads to the equation
$$
\log{(t-1)\over (t+1)}-{2t \over (t^{2}-1)}
={(x-y)\over 8}
\eqno (3.7a)
$$
which can be cast in the form
$$
{(t-1)\over (t+1)}{\rm e}^{2t \over (1-t^{2})}
={\rm e}^{(x-y) \over 8}.
\eqno (3.7b)
$$
This suggests defining
$$
w \equiv {(t-1)\over (t+1)}
\eqno (3.8)
$$
so that one first has to solve the transcendental equation
$$
w{\rm e}^{(w^{2}-1)\over 2w}={\rm e}^{(x-y)\over 8}
\eqno (3.9)
$$
to obtain $w=w(x-y)$, from which one gets
$$
t={(1+w)\over (1-w)}=t(x-y).
\eqno (3.10)
$$
On denoting by $g(w)$ the left-hand side of Eq. (3.9), one finds that,
in the plane $(w,g(w))$, the right-hand side of Eq. (3.9) is a line
parallel to the $w$-axis, which intersects $g(w)$ at no more than
one point for each value of $x-y$. For
example, when $w=1$, $g(w)$ intersects the line taking the constant value
$1$, for which $x-y=0$. The function
$$
g:w \rightarrow g(w)=w{\rm e}^{(w^{2}-1)\over 2w}
$$
is asymmetric and has the limiting behaviour described by
$$
\lim_{w \to 0^{-}}g(w)=-\infty \; \; \;
\lim_{w \to 0^{+}}g(w)=0
\eqno (3.11)
$$
$$
\lim_{w \to -\infty}g(w)=0 \; \; \; 
\lim_{w \to +\infty}g(w)=\infty .
\eqno (3.12)
$$
Thus, in the lower half-plane, $g$ has an horizontal asymptote given
by the $w$-axis, and a vertical asymptote given by the line $w=0$,
while it has no asymptotes in the upper half-plane, since
$$
\lim_{w \to \infty}{g(w)\over w}=\infty
$$
in addition to (3.12). The first derivative of $g$ reads
$$
g'(w)={(w+1)^{2}\over 2w}{\rm e}^{(w^{2}-1)\over 2w}.
\eqno (3.13)
$$
One therefore has $g'(w)>0$ for all $w>0$, and $g'(w)<0$ for all
$w \in (-\infty,0) - \left \{ -1 \right \}$, and $g$ is monotonically
decreasing for negative $w$ and monotonically increasing for positive $w$.
The point $w=-1$, at which $g'(w)$ vanishes, is neither a maximum nor 
a minimum point, because 
$$
g''(w)=\left({1\over 4w^{3}}+{1\over 2w}+1+{w\over 4}\right)
{\rm e}^{(w^{2}-1)\over 2w}
\eqno (3.14)
$$
$$
g'''(w)=\left({1\over 8w^{5}}-{3\over 4w^{4}}
+{3\over 8 w^{3}}+{3\over 8w}+{3\over 4}+{w\over 8}\right)
{\rm e}^{(w^{2}-1)\over 2w}.
\eqno (3.15)
$$
These formulae imply that $g''(-1)=0$ but $g'''(-1)=-1 \not = 0$,
and hence $w=-1$ yields a flex of $g(w)$.

In the $(x,y)$ variables, the operator ${\cal L}_{m,n}$ reads therefore
$$
{\cal L}_{m,n}=f(x,y){\partial^{2}\over \partial x \partial y}
+g(x,y){\partial \over \partial x}+h(x,y){\partial \over \partial y}
+n^{2}-m^{2}
\eqno (3.16)
$$
where, exploiting the formulae
$$
{\partial x \over \partial q}={64 \sqrt{2}\over (t-1)^{2}}
\eqno (3.2b)
$$
$$
{\partial y \over \partial q}={64 \sqrt{2}\over (t+1)^{2}}
\eqno (3.3b)
$$
one finds
$$ \eqalignno{
f(x,y)&=-(2\sqrt{2}+32q)\left({\partial x \over \partial q}
+{\partial y \over \partial q}\right)
+8q^{2}{\partial x \over \partial q}{\partial y \over \partial q}
+128 \cr
&=256 \left[1-{2t^{2}(t^{2}+1)\over (t-1)^{2}(t+1)^{2}}\right]
&(3.17)\cr}
$$
$$ 
g(x,y)=4(n+1)q{\partial x \over \partial q}-16n 
=16 \left[1+{2(n+1)\over (t-1)}\right]
\eqno (3.18)
$$
$$ 
h(x,y)=4(n+1)q{\partial y \over \partial q}-16n 
=16 \left[1-{2(n+1)\over (t+1)}\right].
\eqno (3.19)
$$
The resulting canonical form of Eq. (2.6) is
$$ \eqalignno{
L[\chi]&=\left({\partial^{2}\over \partial x \partial y}
+a(x,y){\partial \over \partial x}
+b(x,y){\partial \over \partial y}+c(x,y)
\right)\chi(x,y) \cr
&={\widetilde H}(x,y)
&(3.20)\cr}
$$
where
$$
a(x,y) \equiv {g(x,y)\over f(x,y)}
={1\over 16}{(1-t)(t+1)^{2}(2n+1+t)\over (t^{4}+4t^{2}-1)}
\eqno (3.21)
$$
$$
b(x,y) \equiv {h(x,y)\over f(x,y)}
={1\over 16}{(t+1)(t-1)^{2}(2n+1-t)\over (t^{4}+4t^{2}-1)}
\eqno (3.22)
$$
$$
c(x,y) \equiv {n^{2}-m^{2}\over f(x,y)}
={(m^{2}-n^{2})\over 256}
{(t-1)^{2}(t+1)^{2}\over (t^{4}+4t^{2}-1)}
\eqno (3.23)
$$
$$
{\widetilde H}(x,y) \equiv {H(x,y)\over f(x,y)}
=-{H(x,y)\over 256}
{(t-1)^{2}(t+1)^{2}\over (t^{4}+4t^{2}-1)}.
\eqno (3.24)
$$
Note that $a(-t)=b(t), b(-t)=a(t), c(-t)=c(t),
{\widetilde H}(-t)={\widetilde H}(t)$.

For an hyperbolic equation in the form (3.20), we can use the Riemann
integral representation of the solution. For this purpose, recall 
from [11] that, on denoting by $L^{\dagger}$ the adjoint of the operator 
$L$ in (3.20), which acts according to
$$
L^{\dagger}[\chi]=\chi_{xy}-(a\chi)_{x}-(b \chi)_{y}+c \chi
\eqno (3.25)
$$
one has to find a `function' $R(x,y;\xi,\eta)$ (actually a kernel)
subject to the following conditions ($(\xi,\eta)$ being the coordinates
of a point $P$ such that characteristics through it intersect a curve
$C$ at points $A$ and $B$, $AP$ being a segment with constant $y$,
and $BP$ being a segment with constant $x$):
\vskip 0.3cm
\noindent
(i) As a function of $x$ and $y$, $R$ satisfies the adjoint equation
$$
L^{\dagger}_{(x,y)}[R]=0
\eqno (3.26)
$$
(ii) $R_{x}=bR$ on $AP$, i.e.
$$
R_{x}(x,y;\xi,\eta)=b(x,\eta)R(x,y;\xi,\eta) \; {\rm on} \;
y=\eta
\eqno (3.27)
$$
and $R_{y}=aR$ on $BP$, i.e.
$$
R_{y}(x,y;\xi,\eta)=a(\xi,y)R(x,y;\xi,\eta) \; {\rm on} \;
x=\xi
\eqno (3.28)
$$
(iii) $R$ equals $1$ at $P$, i.e.
$$
R(\xi,\eta;\xi,\eta)=1.
\eqno (3.29)
$$
It is then possible to express the solution of Eq. (3.20) in the form
$$ \eqalignno{
\chi(P)&={1\over 2}[\chi(A)R(A)+\chi(B)R(B)] \cr
&+\int_{AB}\left( \left[{R\over 2}\chi_{x}
+\left(bR-{1\over 2}R_{x}\right)\chi \right]dx \right . \cr
& \left . - \left[{R\over 2}\chi_{y}
+\left(aR-{1\over 2}R_{y}\right)\chi \right]dy \right) \cr
&+ \int \int_{\Omega}R(x,y;\xi,\eta){\widetilde H}(x,y)dx dy
&(3.30)\cr}
$$
where $\Omega$ is a domain with boundary.

Note that Eqs. (3.27) and (3.28) are ordinary differential equations
for the Riemann function $R(x,y;\xi,\eta)$ along the characteristics
parallel to the coordinate axes. By virtue of (3.29), 
their integration yields
$$
R(x,\eta;\xi,\eta)={\rm exp} \int_{\xi}^{x}b(\lambda,\eta)d\lambda
\eqno (3.31)
$$
$$
R(\xi,y;\xi,\eta)={\rm exp} \int_{\eta}^{y}a(\lambda,\xi)d\lambda
\eqno (3.32)
$$
which are the values of $R$ along the characteristics through $P$.
Equation (3.30) yields instead the solution of Eq. (3.20) for arbitrary
initial values given along an arbitrary non-characteristic curve $C$,
by means of a solution $R$ of the adjoint equation (3.26) which depends 
on $x,y$ and two parameters $\xi,\eta$. Unlike $\chi$, the Riemann
function $R$ solves a characteristic initial-value problem.
\vskip 0.3cm
\leftline {\bf 4. Goursat problem for the Riemann function}
\vskip 0.3cm
\noindent
By fully exploiting the reduction to canonical form of Eq. (2.6)
we have considered novel features with respect to the analysis in
[6, 8], because the Riemann formula (3.30) contains also the integral
along the piece of curve $C$ from $A$ to $B$, and the term
${1\over 2}[\chi(A)R(A)+\chi(B)R(B)]$. This representation of the 
solution might be more appropriate for the numerical purposes
considered in [7], but the task of finding the Riemann function $R$
remains extremely difficult. One can however use approximate methods
for solving Eq. (3.26). For this purpose, we first point out that,
by virtue of Eq. (3.25), Eq. (3.26) is a canonical hyperbolic equation
of the form
$$
\left({\partial^{2}\over \partial x \partial y}
+A{\partial \over \partial x}+B{\partial \over \partial y}
+C \right)R(x,y;\xi,\eta)=0
\eqno (4.1)
$$
where
$$
A \equiv -a
\eqno (4.2)
$$
$$
B \equiv -b
\eqno (4.3)
$$
$$
C \equiv c-a_{x}-b_{y}.
\eqno (4.4)
$$
Thus, on defining
$$
U \equiv R
\eqno (4.5)
$$
$$
V \equiv R_{x}+BR
\eqno (4.6)
$$
the equation (4.1) for the Riemann function is equivalent to the
hyperbolic canonical system [11]
$$
U_{x}=f_{1}(x,y)U+f_{2}(x,y)V
\eqno (4.7)
$$
$$
V_{y}=g_{1}(x,y)U+g_{2}(x,y)V
\eqno (4.8)
$$
where
$$
f_{1} \equiv -B=b
\eqno (4.9)
$$
$$
f_{2}=1
\eqno (4.10)
$$
$$
g_{1} \equiv AB-C+B_{y}=ab-c+a_{x}
\eqno (4.11)
$$
$$
g_{2} \equiv -A=a.
\eqno (4.12)
$$
For the system described by Eqs. (4.7) and (4.8) with boundary data
(3.31) and (3.32) an existence and uniqueness theorem holds (see [11]
for the Lipschitz conditions on boundary data), and we can therefore
exploit the finite differences method to find approximate solutions
for the Riemann function $R(x,y;\xi,\eta)$, and eventually $\chi(P)$
with the help of the integral representation (3.30).
\vskip 0.3cm
\leftline {\bf 5. Concluding remarks}
\vskip 0.3cm
\noindent
The inverses of hyperbolic operators [12] and the Cauchy problem for
hyperbolic equations with polynomial coefficients [13] have always been
the object of intensive investigation in the mathematical literature.
We have here considered the application of such issues to axisymmetric
black hole collisions at the speed of light, relying on the work in
[5--8]. We have pointed out that, for the inhomogeneous equations (2.6)
occurring in the perturbative analysis, the task of inverting the 
operator (2.7) can be accomplished with the help of the Riemann integral
representation (3.30), after solving Eq. (4.1) for the Riemann function.
One has then to solve a characteristic initial-value problem for a
homogeneous hyperbolic equation in canonical form in two independent
variables, for which we have developed formulae to be used for the
numerical solution with the help of a finite differences scheme.
For this purpose one studies the canonical system (cf (4.7) and (4.8))
$$
U_{x}=F(x,y,U,V)
\eqno (5.1)
$$
$$
V_{y}=G(x,y,U,V)
\eqno (5.2)
$$
in the rectangle ${\cal R} \equiv \left \{ x,y:
x \in [x_{0},x_{0}+a], y \in [y_{0},y_{0}+b] \right \}$ with known
values of $U$ on the vertical side $AD$ where $x=x_{0}$, and known
values of $V$ on the horizontal side $AB$ where $y=y_{0}$. The 
segments $AB$ and $AD$ are then divided into $m$ and $n$ equal parts,
respectively. On setting ${a\over m} \equiv h$ and 
${b \over n} \equiv k$, the original differential equations become
equations relating values of $U$ and $V$ at three intersection points
of the resulting lattice, i.e.
$$
{U(P_{r,s+1})-U(P_{rs})\over h}=F
\eqno (5.3a)
$$
$$
{V(P_{r+1,s})-V(P_{rs})\over k}=G.
\eqno (5.4a)
$$
It is now convenient to set $U_{rs} \equiv U(P_{rs}), 
V_{rs} \equiv V(P_{rs})$, so that these equations read
$$
U_{r,s+1}=U_{rs}+hF(P_{rs},U_{rs},V_{rs})
\eqno (5.3b)
$$
$$
V_{r+1,s}=V_{rs}+kG(P_{rs},U_{rs},V_{rs}).
\eqno (5.4b)
$$
Thus, if both $U$ and $V$ are known at $P_{rs}$, one can evaluate
$U$ at $P_{r,s+1}$ and $V$ at $P_{r+1,s}$. The evaluation at
subsequent intersection points of the lattice goes on along horizontal
or vertical segments. In the former case, the resulting algorithm is
$$
U_{rs}=U_{r0}+h \sum_{i=1}^{s-1}F(P_{ri},U_{ri},V_{ri})
\eqno (5.5)
$$
$$
V_{rs}=V_{r-1,s}+kG(P_{r-1,s},U_{r-1,s},V_{r-1,s})
\eqno (5.6)
$$
while in the latter case one obtains the algorithm expressed by the
equations
$$
V_{rs}=V_{0s}+\sum_{i=1}^{r-1}G(P_{is},U_{is},V_{is})
\eqno (5.7)
$$
$$
U_{rs}=U_{r,s-1}+hF(P_{r,s-1},U_{r,s-1},V_{r,s-1}).
\eqno (5.8)
$$
Stability of such solutions is closely linked with the geometry of the
associated characteristics, and the criteria to be fulfilled are
studied in section 13.2 of [14] (stability depends crucially on whether
or not ${h\over k} \leq 1$).

To sum up, one solves numerically Eq. (3.9) for $w=w(x,y)=w(x-y)$,
from which one gets $t(x-y)$ with the help of (3.10), which is a
fractional linear transformation.
This yields $a,b,c$ and $\widetilde H$ as functions of $(x,y)$
according to (3.21)--(3.24), and hence $A,B$ and $C$ in the equation
for the Riemann function are obtained according to (4.2)--(4.4), where
derivatives with respect to $x$ and $y$ are evaluated numerically.
Eventually, the system given by (4.7) and (4.8) is solved according
to the finite-differences scheme of the present section, with
$$
F=f_{1}U+f_{2}V=f_{1}R+f_{2}(R_{x}+BR)
\eqno (5.9)
$$
$$
G=g_{1}U+g_{2}V=g_{1}R+g_{2}(R_{x}+BR).
\eqno (5.10)
$$
Once the Riemann function $R=U$ is obtained with the desired accuracy,
numerical evaluation of the integral (3.30) yields $\chi(P)$, and
$\chi(q,r)$ is obtained upon using Eqs. (3.5) and (3.6) for the
characteristic coordinates.
Our steps are conceptually desirable since they rely on well
established techniques for the solution of hyperbolic equations
in two independent variables [11, 14], and provide a viable 
alternative to the numerical analysis performed in [7], because
all functions should be evaluated numerically. Our method is not
obviously more powerful than the one used in [5--8], but is well
suited for a systematic and lengthy numerical analysis, while its
analytic side provides an interesting alternative for the evaluation
of Green functions both in black hole physics and in other problems
where hyperbolic operators with variable coefficients might occur.
This task remains very important because a strong production of
gravitational radiation is mainly expected in the extreme events studied
in [5--8] and which motivated our paper. Any viable way of looking 
at mathematical and numerical aspects of the problem is therefore
of physical interest for research planned in the years to come [15].
\vskip 0.3cm
\leftline {\bf Acknowledgments}
\vskip 0.3cm
\noindent
The work of G Esposito has been partially supported by the
National project {\sl SINTESI 2000}.
\vskip 1cm
\leftline {\bf References}
\vskip 0.3cm
\noindent
\item {[1]}
DeWitt B S 1965 {\it Dynamical Theory of Groups and Fields}
(New York: Gordon and Breach)
\item {[2]}
DeWitt B S 1984 in {\it Relativity, Groups and Topology II} eds 
B S DeWitt and R Stora (Amsterdam: North--Holland)
\item {[3]}
Esposito G 2001 {\it Quantum Gravity in Four Dimensions}
(New York: Nova Science)
\item {[4]}
Cartier P and DeWitt--Morette C 2000 {\it J. Math. Phys.}
{\bf 41} 4154
\item {[5]}
D'Eath P D and Payne P N 1992 {\it Phys. Rev.} D {\bf 46} 658
\item {[6]}
D'Eath P D and Payne P N 1992 {\it Phys. Rev.} D {\bf 46} 675
\item {[7]}
D'Eath P D and Payne P N 1992 {\it Phys. Rev.} D {\bf 46} 694
\item {[8]}
D'Eath P D 1996 {\it Black Holes: Gravitational Interactions}
(Oxford: Clarendon Press)
\item {[9]}
Esposito G and Stornaiolo C 2000 {\it Class. Quantum Grav.}
{\bf 17} 1989
\item {[10]}
Esposito G and Stornaiolo C 2000 {\it Found. Phys. Lett.}
{\bf 13} 279
\item {[11]}
Courant R and Hilbert D 1961 {\it Methods of Mathematical
Physics. II. Partial Differential Equations} 
(New York: Interscience)
\item {[12]}
Leray J 1953 {\it Hy\-per\-bolic Dif\-fer\-en\-tial Equa\-tions}
(Prince\-ton Uni\-ver\-sity Press)
\item {[13]}
Leray J 1956 {\it C. R. Acad. Sci. Paris} {\bf 242} 953
\item {[14]}
Garabedian P R 1964 {\it Partial Differential Equations}
(New York: Chelsea)
\item {[15]}
Allen B and Ottewill A (2000) {\it Gen. Rel. Grav.} 
{\bf 32} 385

\bye